\documentclass[prl,twocolumn,superscriptaddress,showpacs,amsmath,amssymb,floatfix]{revtex4-1}
\usepackage{graphicx,color}
\usepackage{epsfig}
\usepackage{bm}

\begin{document}

\title{Thermopower with broken time-reversal symmetry}
\author{Keiji Saito}
\affiliation{Department of Physics, Graduate School of Science,
University of Tokyo, Tokyo 113-0033, Japan}
\author{Giuliano Benenti}
\affiliation{CNISM, CNR-INFM \& Center for Nonlinear and Complex Systems,
Universit\`a degli Studi dell'Insubria, Via Valleggio 11, 22100 Como, Italy}
\affiliation{Istituto Nazionale di Fisica Nucleare, Sezione di Milano,
via Celoria 16, 20133 Milano, Italy}
\author{Giulio Casati}
\affiliation{CNISM, CNR-INFM \& Center for Nonlinear and Complex Systems,
Universit\`a degli Studi dell'Insubria, Via Valleggio 11, 22100 Como, Italy}
\affiliation{Istituto Nazionale di Fisica Nucleare, Sezione di Milano,
via Celoria 16, 20133 Milano, Italy}
\affiliation{Centre for Quantum Technologies,
National University of Singapore, Singapore 117543}
\author{Toma\v z Prosen}
\affiliation{Department of Physics, Faculty of Mathematics and Physics,
University of Ljubljana, Ljubljana, Slovenia}

\date{\today}


\begin{abstract}
We show that when non-unitary noise effects are taken into account 
the thermopower is in general asymmetric under magnetic field 
reversal, even for non-interacting systems. 
Our findings are illustrated in the example of a three-dot ring 
structure pierced by an Aharonov-Bohm flux. 
\end{abstract}
\maketitle

In power generation and refrigeration by means of thermal engines,
efficiency plays a basic theoretical and practical role.
The Carnot bound on efficiency lies at the foundations of thermodynamics:
for a heat engine functioning between hot and cold reservoirs at
temperatures $T_h$ and $T_c$, the efficiency $\eta$,  defined as 
the ratio of the output power over the heat extracted per unit time 
from the high temperature reservoir, is upper bounded by the 
Carnot efficiency $\eta_C$: $\eta  \leq \eta_C = 1-T_c/T_h$.

For systems with time-reversal symmetry, thermoelectric power generation 
and refrigeration is governed, within linear response, by a single parameter, 
the dimensionless figure of merit
$ZT=(\sigma S^2/\kappa) T$,
where $\sigma$ is the electric conductivity, $S$ is the
thermopower (Seebeck coefficient),
$\kappa$ is the thermal conductivity, and $T\approx T_h\approx T_c$ is the
temperature.
The maximum efficiency is given by
\begin{equation}
\eta_{\rm max}=
\eta_C\,
\frac{\sqrt{ZT+1}-1}{\sqrt{ZT+1}+1}.
\label{etamaxB0}
\end{equation}
Thermodynamics only imposes $ZT\ge 0$ and 
the Carnot limit is reached when $ZT\to\infty$.

On the other hand, we have recently shown~\cite{ZTmag} 
that for systems with broken time-reversal symmetry the efficiency depends on two parameters: 
a ``figure of merit'' and an asymmetry parameter. In contrast to the time-symmetric case, the figure of merit is
bounded from above; yet the Carnot efficiency can be reached
at lower and lower values of the figure of merit as the asymmetry parameter increases.  
According to the expression for the efficiency, large asymmetry of the thermopower can be responsible
for highly non-trivial effects \cite{ZTmag}, and potentially can be a useful tuning parameter to control thermoelectric
efficiency of the material. Hence, finding general conditions for asymmetry of the thermopower is of general interest both from practical and purely fundamental point of view.

If time-reversal symmetry is broken, e.g. by means of a 
magnetic field ${\bm B}$, then one does not expect the Seebeck 
coefficient to be in general symmetric with respect to the magnetic field. 
Yet for the particular case of non-interacting systems
one has $S({\bm B})=S(-{\bm B})$ as a consequence
of the symmetry properties of the scattering matrix~\cite{datta}.
Even though this constraint does not apply when interactions or 
inelastic scattering are taken into account, and even though there are no
general results imposing the symmetry of the Seebeck coefficient,
the latter has always been found to be an even function of the magnetic
field in purely metallic two-terminal mesoscopic systems~\cite{vanlangen}.
On the other hand, 
Andreev interferometer experiments~\cite{chandrasekhar}
and recent theoretical studies indicate that systems
in contact with a superconductor~\cite{jacquod} or with a 
heat bath~\cite{imry} can exhibit non-symmetric thermopower.
However, accurate numerical simulations of various models 
of two-terminal purely Hamiltonian \textit{interacting} dynamical systems,
which violate time-reversal symmetry, such as a 
two-dimensional anisotropic and inhomogeneous system 
of interacting particles in a perpendicular magnetic field~\cite{carlos}, 
systematically failed to find a non-symmetric thermopower, 
$S({\bm B})\ne S(-{\bm B})$.
Therefore, it is a remains a completely open and interesting 
problem to understand what requirements must 
be fulfilled in order to \textit{actually} lead to a thermopower which is
asymmetric in the magnetic field.

In this Letter we show that the thermopower is in general asymmetric 
when non-unitary noise is added to the system, even though the system is non-interacting. 
Indeed, in the non-interacting 
case the symmetry of the thermopower is a consequence
of the unitarity of the scattering matrix, which is broken when 
noise is added. A very convenient way to introduce noise is by means 
of a third terminal, whose parameters (temperature and chemical potential)
are chosen self-consistently so that there is no average flux of 
particles and heat between the terminal and the system.
In mesoscopic physics, such third terminal, or ``conceptual probe''
is commonly used to simulate phase-breaking processes in 
partially coherent quantum transport, since it 
introduces phase-relaxation without energy damping~\cite{buttiker}. 
We also show that, as a consequence of the asymmetry of the Seebeck coefficient,
a weak magnetic field generally improves either
the efficiency of thermoelectric power generation or of refrigeration,
the efficiencies of the two processes being no longer equal when 
a magnetic field is added. 
Our findings are illustrated by the 
example of a realistic, asymmetric three-dot ring structure 
pierced by an Aharonov-Bohm flux. 
A main advantage of this model is that it can be analyzed exactly,
without resorting to approximations.

\begin{figure}
\begin{center}
\includegraphics[width=7.0cm]{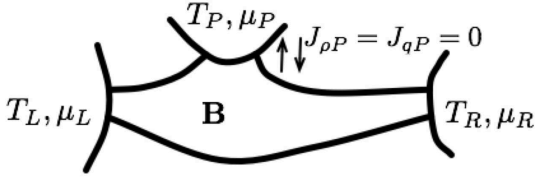}
\caption{Schematic drawing of the model. The third (probe) 
reservoir introduces non-unitary noise.}
\label{fig:scheme}
\end{center}
\end{figure}

\emph{General setup.}
The model we consider is sketched in Fig.~\ref{fig:scheme}.
A system is in contact with left ($L$) and right ($R$) reservoirs
(terminals) at 
temperatures $T_L=T+\Delta T$, $T_R=T$ 
(without loss of generality, we assume $T_L>T_R$) and 
chemical potentials
$\mu_L=\mu+\Delta \mu$, $\mu_R=\mu$. 
Both electric and heat currents flow along the horizontal axis. 
Non-unitary noise effects are 
simulated by means of a third (probe) reservoir ($P$) at temperature 
$T_P=T+\Delta T_P$ and chemical potential
$\mu_P=\mu+\Delta\mu_P$. 
Let $J_{\rho\, k}$ and $J_{E\, k}$ denote the particle and energy
currents from the $k$-th reservoir ($k=L,R,P$) into the system,
with the steady-state constraints of charge and energy conservation:
$\sum_k J_{\rho \, k}=0$, $\sum_k J_{E\, k}=0$. The sum of the entropy
production rates at the reservoirs reads
$\dot{S}=\sum_k(J_{E\, k}-\mu_k J_{\rho\, k})/T_k$. 
Within linear response, 
$\dot{S}= {\bm J} \cdot {\bm X} \equiv
\sum_{i=1}^4 J_i X_i$, where we have defined 
the 4-dimensional vectors ${\bm J}$ and ${\bm X}$:
\begin{equation}
 {\bm J}= (eJ_{\rho L},J_{q L},eJ_{\rho P},J_{q P}),
\end{equation}
\begin{equation}
 {\bm X}=\left(\frac{\Delta \mu}{eT},\frac{\Delta T}{T^2},
\frac{\Delta \mu_P}{e T},\frac{\Delta T_P}{T^2}\right),
\end{equation}
and where the heat currents $J_{q\,k}\equiv J_{E\,k}-\mu J_{\rho\,k}$
and $e$ is the electron charge. The equation connecting 
the fluxes $J_i$ and the thermodynamic forces $X_i$ within
linear irreversible thermodynamics is~\cite{callen} 
\begin{equation}
{\bm J}={\bm L}{\bm X},
\label{eq:onsager4}
\end{equation}
where ${\bm L}$ is a $4\times 4$ Onsager matrix.

The probe reservoir is adjusted in such a way that 
$J_3=J_4=0$, that is, the net particle and heat
flow from the probe into the system vanishes. 
It is convenient to write Eq.~(\ref{eq:onsager4}) in the 
block matrix form
\begin{equation}
\left(
\begin{array}{c}
{\bm J_\alpha}\\
{\bm J_\beta}
\end{array}
\right)
=
\left(
\begin{array}{cc}
{\bm L_{\alpha\alpha}} & 
{\bm L_{\alpha\beta}} \\ 
{\bm L_{\beta\alpha}} & 
{\bm L_{\beta\beta}} 
\end{array}
\right)
\left(
\begin{array}{c}
{\bm X_\alpha}\\
{\bm X_\beta}
\end{array}
\right),
\end{equation}
where ${\bm \alpha}$ stands for $(1,2)$
and ${\bm \beta}$ for $(3,4)$.
The self-consistency condition ${\bm J_\beta}=(J_3,J_4)=0$ implies 
${\bm X_\beta}=-{\bm L_{\beta\beta}}^{-1}{\bm L_{\beta\alpha}}
{\bm X_\alpha}$, so that
\begin{equation}
{\bm J_\alpha}=
{\bm L}^\prime
{\bm X_\alpha},
\quad
{\bm L}^\prime
\equiv
 {\bm L_{\alpha\alpha}}-
{\bm L_{\alpha\beta}} 
{\bm L_{\beta\beta}}^{-1} 
{\bm L_{\beta\alpha}}. 
\label{eq:reduction}
\end{equation} 
The problem has then been reduced to two coupled fluxes:
\begin{equation}
\left(
\begin{array}{c}
J_1\\
J_2
\end{array}
\right)
=
\left(
\begin{array}{cc}
L_{11}^\prime & L_{12}^\prime \\
L_{21}^\prime & L_{22}^\prime 
\end{array}
\right)
\left(
\begin{array}{c}
X_1\\
X_2
\end{array}
\right),
\end{equation}
where the reduced $2\times 2$ Onsager matrix
matrix ${\bm L}^\prime$ 
fulfills the Onsager-Casimir relations
\begin{equation}
L_{ij}^\prime({\bm B})=L_{ji}^\prime(-{\bm B}),
\quad (i,j=1,2).
\end{equation}
We would like to draw the reader's attention to the fact that the 
matrix ${\bm L}^\prime$ is the Onsager matrix for
two-terminal noisy transport, with noise modeled 
by means of a self-consistent reservoir. In particular, the Seebeck 
and the Peltier coefficients are given by 
$S=L_{12}^\prime/(eTL_{11}^\prime)$ and 
$\Pi=L_{21}^\prime/(eL_{11}^\prime)$. The thermopower is asymmetric 
when $L_{12}^\prime({\bm B})\ne L_{21}^\prime({\bm B})$,
i.e. $\Pi\ne S T$.

A key point is that, since $J_3=J_4=0$, 
$J_1$ is the charge current from left to right reservoir
and the heat is extracted  from (for power generation) or dissipated to
(for refrigeration) the left (or right) reservoir only. Therefore, 
we can apply the analysis developed in Ref.~\cite{ZTmag}. 
In particular, the efficiency
depends on the asymmetry parameter $x$ and 
on the ``figure of merit'' parameter $y$: 
\begin{equation}
x=\frac{L_{12}^\prime}{L_{21}^\prime},
\quad
y=\frac{L_{12}^\prime L_{21}^\prime}{{\rm det}{\bm L}^\prime}.
\end{equation}
For power generation ($J_2>0$ and output power 
$\omega=-J_1 \Delta\mu=-J_1eTX_1>0$) the efficiency 
$\eta=\omega/J_2$ has a maximum value
\begin{equation}
\eta_{\rm max}=\eta_C \,x\,\frac{\sqrt{y+1}-1}{\sqrt{y+1}+1},
\label{eq:etapower}
\end{equation}
while for refrigeration ($J_2<0$, $\omega<0$) the maximum 
of the efficiency $\eta^{(r)}=J_2/\omega$ is
\begin{equation}
\eta_{\rm max}^{(r)}=\eta_C \,\frac{1}{x}\,\frac{\sqrt{y+1}-1}{\sqrt{y+1}+1}.
\label{eq:etarefrigeration}
\end{equation}

\emph{Non-interacting systems.}
Exact calculation of thermopower and efficiencies is possible 
for non-interacting models by means of the Landauer-B\"uttiker approach.
We start from the bilinear Hamiltonian
$H=H_S+H_R+H_C$, where the different terms correspond,
respectively, to the
nanoscale electronic system, the reservoirs, and the reservoir-system
coupling. The tight-binding $N$-site system Hamiltonian reads
\begin{equation}
H_S=\sum_{n,n'=1}^N H_{nn'} c_n^\dagger c_n',
\end{equation}
where $c_n$ and $c_n^\dagger$ are 
fermionic annihilation and creation operators. The reservoirs are 
modeled as ideal fermi gases:
$H_R=\sum_{k,q} E_q c_{kq}^\dagger c_{kq}$, where $c_{kq}^\dagger$
creates an electron in the state $q$ in the $k$-th reservoir. 
The coupling (tunneling) Hamiltonian
\begin{equation}
H_C=\sum_{k,q} (t_{kq}c_{kq}^\dagger c_{i_k} +
t_{kq}^* c_{kq} c_{i_k}^\dagger)
\end{equation}
establishes the contact between site $i_k$ and reservoir 
$k$~\cite{footnote}.

The charge and heat currents from the left terminal (reservoir) 
are given by~\cite{sivan}
\begin{equation}
J_1=\frac{e}{h}\int_{-\infty}^{\infty}
dE \sum_k[T_{kL}(E)f_L(E)-T_{Lk}(E)f_k(E)],
\end{equation}
\begin{equation}
J_2=\frac{1}{h}\int_{-\infty}^{\infty}
dE (E-\mu_L) \sum_k[T_{kL}(E)f_L(E)-T_{Lk}(E)f_k(E)],
\end{equation}
where $f_k(E)=\{\exp[(E-\mu_k)/k_B T_k]+1]^{-1}$
is the Fermi function and 
$T_{kl}$ is the transmission probability from 
terminal $l$ to terminal $k$. Analogous expressions can be 
written for $J_3$ and $J_4$, provided the terminal $L$
is substituted by $P$.

The Onsager coefficients $L_{ij}$ are obtained from the 
linear response expansion of the currents $J_i$. We have
\begin{equation}
L_{11}=\frac{e^2}{h}\int_{-\infty}^\infty
dE \sum_{k\ne L} T_{Lk}(E) F(E),
\label{eq:L11}
\end{equation}
\begin{equation}
L_{12}=L_{21}=
\frac{e}{h}\int_{-\infty}^\infty
dE (E-\mu) \sum_{k\ne L} T_{Lk}(E) F(E),
\end{equation}
\begin{equation}
L_{22}=\frac{1}{h}\int_{-\infty}^\infty
dE (E-\mu)^2\sum_{k\ne L} T_{Lk}(E) F(E),
\end{equation}
where $F(E)\equiv -T f^\prime(E)=1/4k_B\cosh^2[(E-\mu)/k_BT]$.
Analogous formulas are obtained for 
$L_{33}$, $L_{34}=L_{43}$, and $L_{44}$, with the 
$P$ terminal used instead of $L$.  
Note that for the non-interacting three-teminal model
$L_{12}$ is still an even function of the 
magnetic field, that is, $L_{12}=L_{21}$. On the other hand 
the symmetry of the off-diagonal 
matrix elements is broken for the reduced Onsager matrix
${\bm L}^\prime$. Indeed, reduction (\ref{eq:reduction})
involves other off-diagonal matrix elements of ${\bm L}$ -- between `left' $(1,2)$ and `probe' $(3,4)$ sectors -- which 
in general are not even functions of an applied magnetic
field. The block ${\bm L_{\alpha\beta}}$ of matrix 
${\bm L}$ is given by
\begin{equation}
{\bm L_{\alpha\beta}}
=-\frac{e^2}{h}\int_{-\infty}^\infty 
dE
\left(
\begin{array}{cc}
1 & \frac{E-\mu}{e}\\
\frac{E-\mu}{e} & \left(\frac{E-\mu}{e}\right)^2
\end{array}
\right) 
T_{LP}(E)F(E), 
\end{equation}
and ${\bm L_{\alpha\beta}}
\ne {\bm L_{\beta\alpha}}$,
since ${\bm L_{\beta\alpha}}$ is obtained from 
${\bm L_{\alpha\beta}}$
after substitution of $T_{LP}$ with $T_{PL}$ and 
in general $T_{LP}\ne T_{PL}$.

The transmission probabilities are given by~\cite{datta}
\begin{equation}
T_{pq}={\rm Tr}[\Gamma_p(E)G(E)\Gamma_q(E)G^\dagger(E)], 
\label{eq:Tpq}
\end{equation}
where the broadening matrices $\Gamma_k$ are defined in terms
of the self-energies $\Sigma_k$: 
$\Gamma_k(E)\equiv i[\Sigma_k(E)-\Sigma_k^\dagger(E)]$ and the 
(retarded) system Green function 
$G(E)\equiv [E-H_S-\sum_k \Sigma_k(E)]^{-1}$. 

\begin{figure}
\begin{center}
\includegraphics[width=6.0cm]{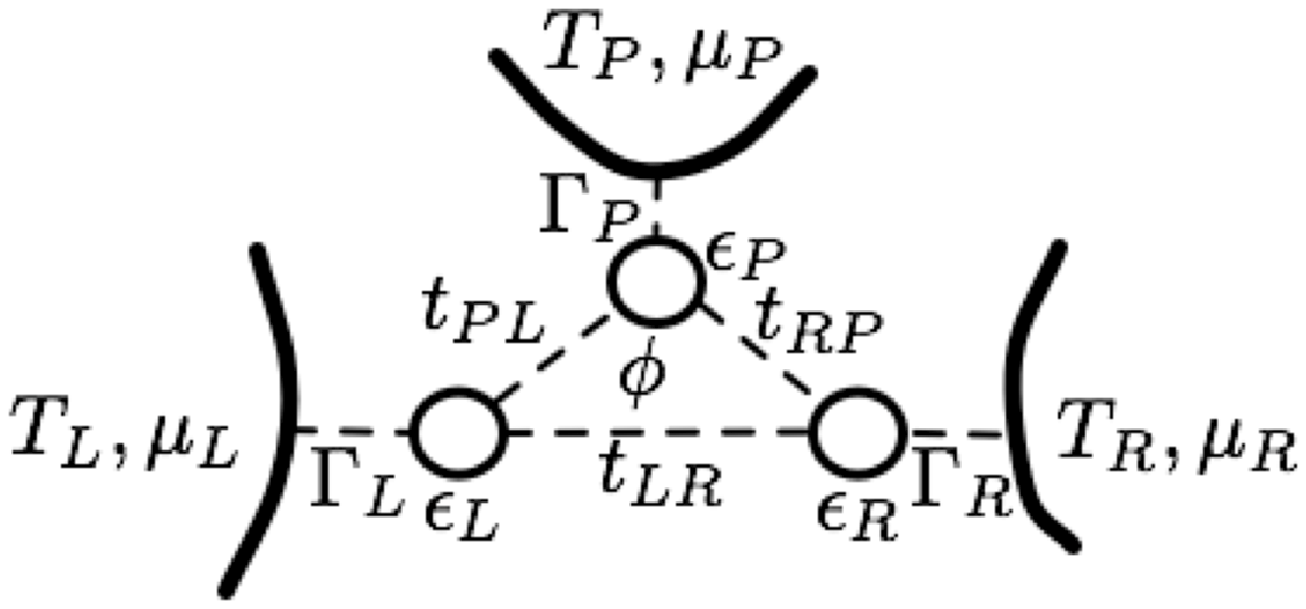}
\caption{Schematic drawing of the three-dot model.} 
\label{fig:3dots}
\end{center}
\end{figure}

\emph{Aharonov-Bohm interferometer.}
As an illustrative, realistic example we consider a three-dot
ring structure pierced by an Aharonov-Bohm flux, with dot $k$
coupled to reservoir $k$, as sketched in Fig.~\ref{fig:3dots}.
The system Hamiltonian reads
\begin{equation}
\begin{array}{c}
H_S=\sum_{k}\epsilon_k c_k^\dagger c_k +
\\
\\
(t_{LR}c_R^\dagger c_L e^{i\phi/3}+
t_{RP}c_P^\dagger c_R e^{i\phi/3}+
t_{PL}c_L^\dagger c_P e^{i\phi/3}+\hbox{H.c.}),
\end{array}
\end{equation}
and the broadening matrices are $\Gamma_k=\gamma_k c_k^\dagger c_k$. 
We apply the Landauer-B\"uttiker approach to this model, 
numerically computing the Onsager coefficients 
following Eqs.~(\ref{eq:L11})-(\ref{eq:Tpq}).

\begin{figure}[!ht]
\begin{center}
\includegraphics[scale=0.45]{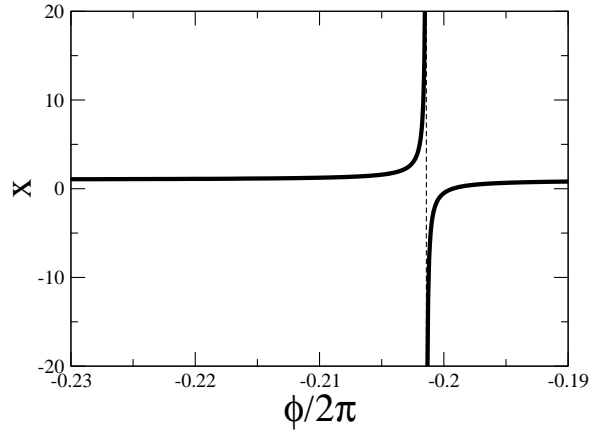}
\caption{Ratio $x$ of the 
off-diagonal matrix elements $L_{12}^\prime$ 
and $L_{21}^\prime$ of the reduced Onsager matrix 
at $T=1$, $\mu=0.3$, 
$\epsilon_L=0$, $\epsilon_R=0.5$, $\epsilon_P=1$,
all hopping terms $t_{pq}=-1$, broadenings $\gamma_k=0.1$ 
independently of energy (wide-band limit). 
Hereafter we set $e=\hbar=k_B=1$.} 
\label{fig:asymmetry}
\end{center}
\end{figure}

As expected, we obtain
asymmetric off-diagonal reduced Onsager matrix elements,
that is $L_{12}^\prime\ne L_{21}^\prime$, as 
far as the Aharonov-Bohm flux $\phi$ is non-vanishing and there is 
anisotropy in the systems, for instance when 
$\epsilon_L\ne\epsilon_R$. 
Since the thermopower is not symmetric with respect to
the magnetic field, i.e. 
$L_{12}^\prime({\bm B})\ne L_{12}^\prime(-{\bm B})= 
L_{21}^\prime({\bm B})$, then in general the ratio 
$x=L_{12}^\prime/L_{21}^\prime\ne 1$.
The asymmetry parameter $x$ can be made arbitrarily 
small when $L_{21}^\prime\to 0$ or arbitrarily large  
when $L_{21}^\prime\to 0$, see for instance Fig.~\ref{fig:asymmetry}.

\emph{Remarks.} 
Large asymmetries \textit{ipso facto} do not imply
large efficiencies. 
For example, in the case of Fig.~\ref{fig:asymmetry} 
when $x$ diverges the figure of merit $y$ and the efficiency
tend to zero.
It is however interesting to compare the efficiencies of power generation and
refrigeration. 
While in the time-symmetric case the two efficiencies coincide,
$\eta_{\rm max}(\phi=0)=\eta_{\rm max}^{(r)}(\phi=0)$,
this is no longer the case when $x\ne1$. 
For small fields $x$ is in general
a linear function of the field, 
$x(\phi)=1+\alpha\phi+O(\phi^2)$,
while $y$ is by construction an even function of the field, 
so that $y(\phi)=y(0)+\beta\phi^2+O(\phi^4)$.
From Eqs.~(\ref{eq:etapower}) and (\ref{eq:etarefrigeration})
we obtain 
$\eta_{\rm max}(\phi)=\eta_{\rm max}(0)[1+\alpha\phi+O(\phi^2))]$
and 
$\eta_{\rm max}^{(r)}(\phi)=\eta_{\rm max}(0)[1-\alpha\phi+O(\phi^2))]$.
Therefore, a small external magnetic field either improves power generation and 
worsens refrigeration or vice-versa, while the average 
efficiency
$\bar{\eta}\equiv 
[\eta_{\rm max}(\phi)+\eta_{\rm max}^{(r)}(\phi)]/2
=\eta_{\rm max}(0)$ 
up to second order corrections.
Due to the Onsager-Casimir relations $x(-\phi)=1/x(\phi)$ 
and therefore by inverting the direction of the magnetic field 
one can improve either power generation or refrigeration.

In conclusion, we have shown that non-unitary noise generally leads 
to a thermopower which is a non-symmetric function of the magnetic field.
Such general result has been illustrated by means of a  
realistic three-dot Aharonov-Bohm interferometer model, 
which appears suitable for experimental investigations by 
means of three-terminal mesoscopic devices. 
The asymmetry of the Seebeck coefficient with respect to 
the magnetic field allows in principle, 
\textit{in the linear response regime},
to obtain a finite power at Carnot efficiency.
Whether this is actually the case remains an interesting open problem. 
An additional interesting open problem is whether noiseless interacting 
systems might exhibit asymmetric thermopower. 

KS was supported by MEXT, Grant Number (23740289),
GB and GC by the
MIUR-PRIN 2008 and by Regione Lombardia, 
and TP by the Grants J1-2208 and P1-0044 of Slovenian Research Agency.


\begin{thebibliography}{0}

\bibitem{ZTmag}
G. Benenti, K. Saito, and G. Casati, 
Phys. Rev. Lett. \textbf{106}, 230602 (2011).

\bibitem{datta}
S. Datta, \textit{Electronic Transport in Mesoscopic Systems}
(Cambridge University Press, 1995).

\bibitem{vanlangen}
S.A. van Langen, P.G. Silvestrov, and C.W.J. Beenakker, 
Superlattices Microstruct. \textbf{23}, 691 (1998); 
S.F. Godijn, S. Moller, H. Buhmann, L.W. Molenkamp, and 
S.A. van Langen, Phys. Rev. Lett. \textbf{82}, 2927 (1999).

\bibitem{chandrasekhar}
J. Eom, C.-J. Chien, and V. Chandrasekhar,
Phys. Rev. Lett. \textbf{81}, 437 (1998).

\bibitem{jacquod}
Ph. Jacquod and R.S. Whitney,
Europhys. Lett. \textbf{91}, 67009 (2010).

\bibitem{imry}
O. Entin-Wohlman, Y. Imry, and A. Aharony, preprint.

\bibitem{carlos}
C. Mej\'{\i}a-Monasterio, private communication.

\bibitem{buttiker}
M. B\"uttiker, IBM J. Res. Developm. \textbf{32}, 63 (1988).

\bibitem{footnote}
To simplify writing we have considered spinless fermions; 
however, the model can be readily extended to include the spin degrees 
of freedom.

\bibitem{sivan}
U. Sivan and Y. Imry,
Phys. Rev. B \textbf{33}, 551 (1986);
P.N. Butcher, J. Phys.: Condens. Matter
\textbf{2}, 4869 (1990).

\bibitem{callen}
H.B. Callen, \textit{Thermodynamics and an Introduction to Thermostatics}
(second edition) (John Wiley \& Sons, New York, 1985).

\end{thebibliography}
\end{document}